# DNN-based workflow for attenuating seismic interference noise and its application to marine towed streamer data from the Northern Viking Graben

Jing Sun[1, 2], Song Hou[3], and Alaa Triki[3]

[1]University of Oslo, Department of Geosciences, Sem Sælands vei 1, Oslo No 0371, Norway. Email: jingsun8803@hotmail.com.

[2]CGG Services (Norway) AS, Lilleakerveien 6A, Box 43 Lilleaker, Oslo No 0216, Norway.

[3]CGG Services (UK) Ltd, Crompton Way, Manor Royal Estate, Crawley, West Sussex RH10 9QN, UK. Email: song.hou@cgg.com; alaa.triki@cgg.com.

## ABSTRACT

To separate seismic interference (SI) noise while ensuring high signal fidelity, we propose a deep neural network (DNN)-based workflow applied to common shot gathers (CSGs). In our design, a small subset of the entire to-be-processed data set is first processed by a conventional algorithm to obtain an estimate of the SI noise (from now on called the SI noise model). By manually blending the SI noise model with SI-free CSGs and a set of simulated random noise, we obtain training inputs for the DNN. The SI-free CSGs can be either real SI-free CSGs from the survey or SI-attenuated CSGs produced in parallel with the SI noise model from the conventional algorithm depending on the specific project. To enhance the DNN's output signal fidelity, adjacent shots on both sides of the to-be-processed shot are used as additional channels of the input. We train the DNN to output the SI noise into one channel and the SI-free shot along with the intact random noise into another. Once trained, the DNN can be applied to the entire data set contaminated by the same types of SI in the training process, producing results efficiently. For demonstration, we applied the proposed DNN-based workflow to 3D seismic field data acquired from the Northern Viking Graben (NVG) of the North Sea, and compared it with a conventional algorithm. The studied area has a challenging SI contamination problem with no sail lines free from SI noise during the acquisition. The comparison shows that the proposed DNN-based workflow outperformed the conventional algorithm in processing quality with less noise residual and better signal preservation. This validates its feasibility and value for real processing projects.

## INTRODUCTION

During acquisition of a seismic survey, unwanted acoustic energy from seismic sources not linked to the survey might be recorded. Such (coherent) SI noise is a significant problem and is harmful to a number of processing operations such as deghosting, demultiple, velocity estimations and amplitude versus offset (AVO) analysis (Gulunay et al., 2004). The angles of incidence for SI noise normally differs both within a sail line and from sail line to sail line within one survey, depending on the relative placement of the external sources to the receivers. Likewise, the amplitudes of SI noise can also vary greatly depending on both the relative distance between the external sources and the receivers as well as the size of the external source(s). As SI noise normally is generated by dedicated sources for seismic exploration, it tends to be well preserved over large distances (Akbulut et al., 1984; Jansen et al., 2013) and may overlap with reflections from sub-surface layers which have significantly lower amplitudes.

For signal separation tasks such as SI attenuation, signal fidelity is a critical metric of the processing quality. It denotes the accuracy of a selected algorithm in preserving the true signal. As SI noise appears intermittently in a field survey while the underlying reflection data remain unknown, quality control (QC) in actual processing projects of SI attenuation is commonly based on visual inspection. To further assess how much SI has been removed after running a SI attenuation algorithm, the root mean square (RMS) amplitude maps in the deep section (where the SI mainly exists, typically after 5.0 s) can be calculated. In addition, a practical way to find signal leakages is to sort data into the channel domain where they are more visible and easier to note. Note that the word "channel" is used as a terminology in both fields of seismic processing and deep learning (DL) with different meanings. In this paper, the word "channel" is always used in the DL lingo, which assigns a multi-dimensional representation to each pixel location as transmitted from its definition in the field of conventional image processing, except in cases when it is used in the seismic term "channel domain" (Sheriff, 2002) as in this paragraph.

SI noise is typically observed at different arrival times in each seismic shot recording. Therefore, existing algorithms for SI noise attenuation often use a strategy of data resorting to obtain a more incoherent distribution of the SI noise in the common receiver or common offset domain. Data transforms, e.g., tau-p transform, are also commonly adopted in an attempt to discriminate the noise via differences in dips/curvature from the underlying signal. After that, a filtering-based denoising, e.g., f-x prediction filters can be applied (Wang et al., 1989; Gulunay and Pattberg, 2001a, b; Guo and Lin, 2003; Kommedal et al., 2007; Elboth et al., 2010). To remove SI noise with different moveout (dip and/or curvature) for a whole survey, different sets of parameters normally need to be manually tested and selected in the application of a conventional algorithm. This process can be laborious. In addition, in some cases, e.g., when SI noise comes from two directions or has similar dip to the underlying seismic reflection data, existing algorithms may perform defectively with noticeable noise residual and signal leakage.

In recent years, encouraged by successful applications in conventional image processing, many applications of DNNs have been made in the field of seismic processing, including SI noise attenuation (Slang, 2019; Sun et al., 2020; Xu et al., 2020). Nevertheless, most studies of seismic data processing have only verified the advantage of DNN-based approaches in saving overall processing time and labor costs, i.e., the processing efficiency. Some studies also showed DNNs' outperformance in terms of processing quality when compared with a single, individually applied, conventional method; but once it comes to competing with a production workflow integrating a series of conventional signal processing methods, DNNs can easily fail. This is the main reason why DNNs have not been widely deployed in actual seismic processing projects. In addition, DNNs can be fooled (Nguyen et al., 2015), and the uncertainty in their performance cannot be easily measured (Ghahramani, 2015).

In this work, we propose a practical DNN-based workflow for separating SI noise whilst preserving the fidelity of the wanted signal on the CSGs. This paper is organized as follows. First, we present the DNN-based workflow with a detailed elaboration of each step. Then, we apply the proposed DNN-based workflow to field data acquired from a marine seismic survey contaminated by various types of SI noise on different sail lines. The studied area was part of the mega NVG survey and was contaminated during the acquisition with no sail line free from SI noise. Therefore, a conventional SI noise attenuation algorithm was applied to produce both the SI noise model and SI-free CSGs for the training and validation of the DNN. After being properly trained, the DNN was applied to unseen field SI-contaminated data and compared with the conventional algorithm. Comparison shows that the proposed DNN-based workflow outperformed the conventional algorithm in terms of processing quality, which demonstrates its feasibility and value for real processing projects. Last but not least, we discuss the impact of the conventional algorithm that provided training data to the DNN and the potential of using the proposed workflow to deal with other types of coherent seismic noise.

## THE DNN-BASED WORKFLOW

We design a supervised deep learning workflow for separating SI noise from seismic data. A schematic diagram is shown in Figure 1. To generate the training and validation pairs, we first process a subset (purple dotted line) of the entire data set (typically a 3D survey) contaminated by SI noise during the field acquisition (the purple ellipse) through a conventional algorithm to obtain the SI noise model (the pink box). Depending on the project, SI-free shots (the orange box) can be obtained via two means. The first option suits surveys that happen to have a few sail lines free from SI noise during the acquisition (orange arrow with dashed line). These data can be directly used for manually blending training input data for the DNN. For another more challenging case where all sail lines of the studied survey are field SI contaminated, we can only adopt the second option, which is to use SI attenuation results produced simultaneously with the SI noise model from a conventional SI attenuation algorithm (orange arrow with solid line).

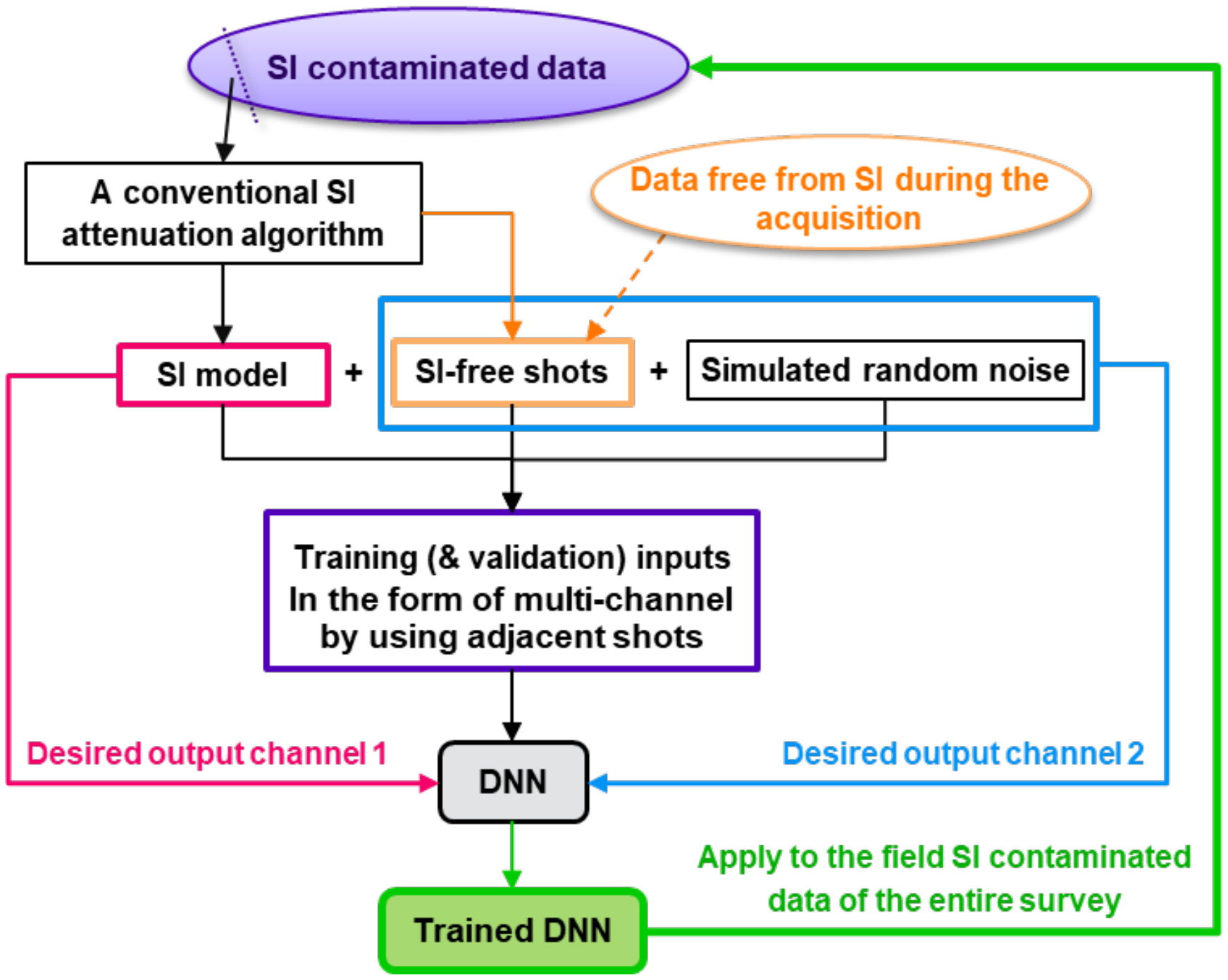

Figure 1: Schematic diagram of the DNN-based workflow proposed for SI noise attenuation on the CSGs.

By manually blending the SI noise model with SI-free CSGs, we can obtain SI-contaminated CSGs with control of the ground truth. The SI noise model prepared for DNN training should cover all the types of SI noise that need to be processed at the DNN application stage. Data augmentation methods (Perez and Wang, 2017; Shorten and Khoshgoftaar, 2019) can always be used for training data preparation, of which the selective methods can be, but are not limited to, random scaling, flipping, shifting, cropping, rotation and tilting. The main purpose here is to create a larger training set so that less SI-contaminated data will need to be run through the conventional physics-based algorithm and the preparation time will be reduced. The main principle of applying such data augmentation methods is to make the new samples look realistic and relevant for the survey of interest. In addition, randomly shuffling the entire data set (for training and validation) is recommended. This avoids similar SI noise types always appearing in regular succession and will benefit the DNN's predictive performance and generalizability.

To enhance the signal fidelity of the DNN's output, one key step is to simulate a set of (can be but not limited to uniform or normal) random noise and inject it into the SI-contaminated CSGs (Sun and Hou, 2022). This blend will be used as the training and validation input data for the DNN (the purple box). Learnable high-level features exist in both seismic signals and coherent noise but not in random noise. Injecting additional random noise into the input can steer the DNN to better focus on learning the high-level features of the data. We tested different intensity groups of the to-be-injected random noise and found they were not that sensitive to the DNN model accuracy as long as they remained visible in the shot gathers but not fully covering the SI noise. Too weak random noise cannot attract the DNN's attention, while too strong random noise will bind the DNN to the covered information. An empirical suggestion is to simulate random noise with an RMS amplitude that varies from 30% to 80% of the RMS amplitude of the field SI contaminated shot.

This is also the range we used in this case study.

In the proposed workflow, the DNN learns to predict two channels in the output. One is the SI noise; the corresponding trajectory is represented as a pink arrow with “Desired output channel 1” in Figure 1. The other is the merge of the SI-free shot and the intact additional random noise (the blue box). The trajectory of this prediction is marked as a blue arrow with “Desired output channel 2” in Figure 1. In this way, the DNN will learn that all the information in the given input should be preserved, albeit to be output to different channels. In other words, the summation of the DNN’s outputs should be exactly equal to the input. From this perspective, predicting the injected random noise along with the SI or the SI-free shot has no theoretical difference. However, as mentioned above, random noise has no learnable high-level features, but only low-level features. After being trained to keep all input information but to send different data components into different channels, the DNN will learn to pass low-level features more heavily into the channel that we choose for random noise due to their commonality in characteristics (Sun and Hou, 2022). In our task, low-level features for SI-free shots are more of a concern, and we thereby predict random noise in this channel.

Another technique adopted in the proposed workflow for enhancing the DNN’s signal fidelity is to use a number of adjacent shots on both sides of the to-be-processed shot as additional channels of the input. This technique is suitable for separating noise that is coherent in the shot domain but reasonably incoherent in the channel domain from the wanted reflection signals. Its effectiveness is demonstrated by Sun et al. (2022) on a shot-domain seismic deblending task based on comparison with the case when an input set of only a single channel is used. The key behind this technique is to allow the DNN to make use of information in both the shot and channel domains: seismic events from consecutive shots are correlated and continuous, whereas the noise (e.g., SI noise, blending noise) tends to be uncorrelated and discontinuous. Once trained, the DNN can be directly applied to the entire field SI-contaminated data of interest. At this stage, there is no need to inject random noise into the SI-contaminated shot records input to the DNN.

## CASE STUDY: A MARINE TOWED STREAMER SURVEY

### Survey and pre-processing

In this section, we demonstrate the effectiveness of the proposed DNN-based workflow on a real processing project of a marine survey conducted in the NVG region of the North Sea, as shown in Figure 2. The NVG survey (Xiao et al., 2018; Latter et al., 2018) was acquired offshore Norway from 2014 to 2016 using BroadSeis (Soubaras et al., 2012) variable-depth streamers in conjunction with a broadband source to maximize the bandwidth. The data studied in this paper were acquired from the eastern region of the work area in a north-south acquisition direction. In Figure 2, the red box represents the survey, while the green box marks the location of the studied data. The actual acquisition geometry is shown in Figure 3. The acquisition configuration had a dual-source array and a sail-line separation of 450 m. There were 12 streamers in total, each 75 m apart and measuring 7950 m in length (636 receivers with a spacing of 12.5 m).

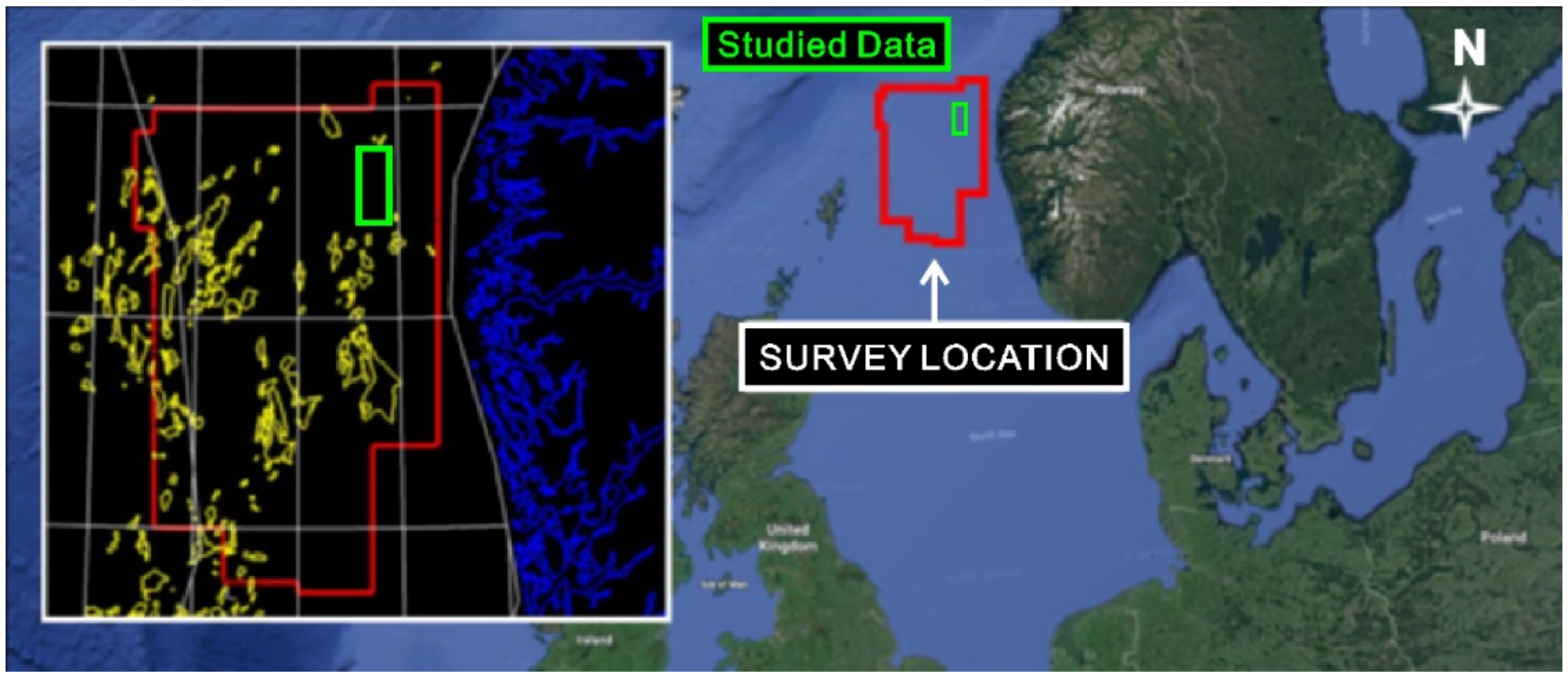

Figure 2: The location of the NVG survey in the northern North Sea (Xiao et al., 2018) as well as the location of the studied data.

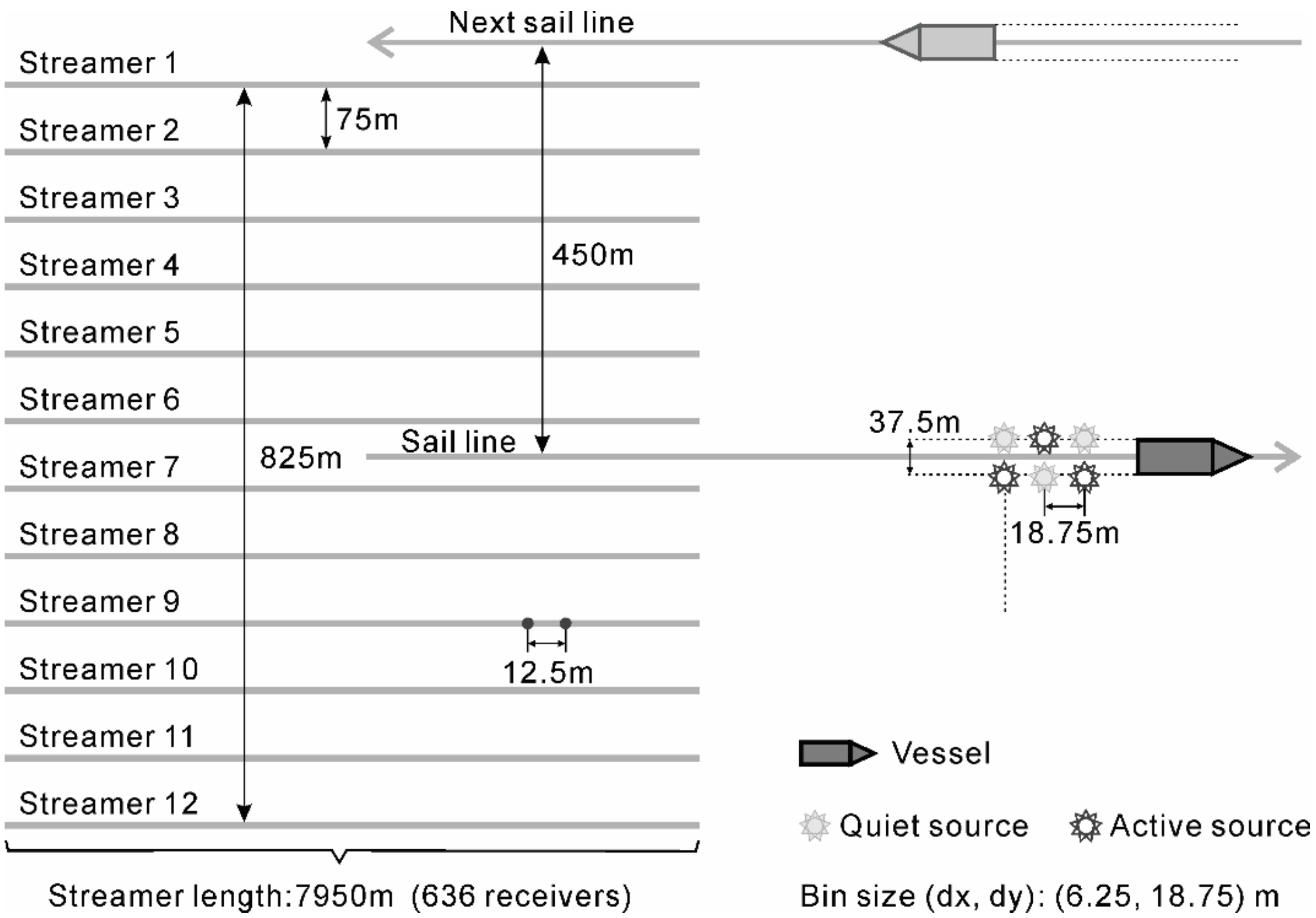

Figure 3: Layout of the towed streamer acquisition survey conducted in the NVG region of the northern North Sea.

The acquisition was conducted during a busy summer season of seismic (and fishing) activities, where both a large exploration survey, as well as several 4D's were acquired simultaneously in a relatively small area. This resulted in SI noise of various characteristics being recorded, including some types that are particularly difficult to remove via conventional means. As shown in Figures 4a and 4b, SI noise coming from two directions was sometimes recorded at the same time. These types of SI noise have high amplitudes and different dips/moveout, and some of them are nearly horizontal, as shown in Figure 4b. In addition, there is SI noise with dip similar to that of the wanted signal (Figure 4c) and SI noise with opposing dip (Figure 4d). Before the attenuation of SI noise, data pre-processing consisted principally of a time-square ($T^2$) spherical divergence gain correction, broadband specific low-cut filter (2.5 Hz), resample to 4 ms sample rate and swell noise attenuation.

**The conventional physics-based SI attenuation algorithms**

Tau-p transforms give good representations of the seismic signals and have been widely used in many processing tasks, including the attenuation of SI noise (Gulunay et al., 2007). A general description of the tau-p-based SI attenuation methodology can be found in Laurain and Elboth (2017): First, we transfer the SI-contaminated shot (2D source-cable gather) into the tau-p domain and sort the transformed data according to p-values; after that, we can apply a random noise attenuation tool to the common-p gathers before the identified noise is sorted back to the tau-p domain, reverse transformed and finally subtracted from the original data. A similar but possibly better algorithm is to replace the regular tau-p transform with the progressive sparse tau-p inversion (Zhang and Wang, 2015), which was first proposed for plane-wave decomposition in the presence of strong spatial aliasing (Wang and Nimsaila, 2014). This is because the regular tau-p transform may suffer from energy leakage among different slowness values, which limits the separability of events with different apparent dips. In contrast, the use of progressive sparse tau-p inversion can provide a more accurate tau-p representation of the input shot gather, resulting in a better SI attenuation result (Zhang and Wang, 2015). This sparse tau-p inversion algorithm has been widely used in real processing projects of SI attenuation. It also produces the SI-free data and the SI noise model we used in DNN training.

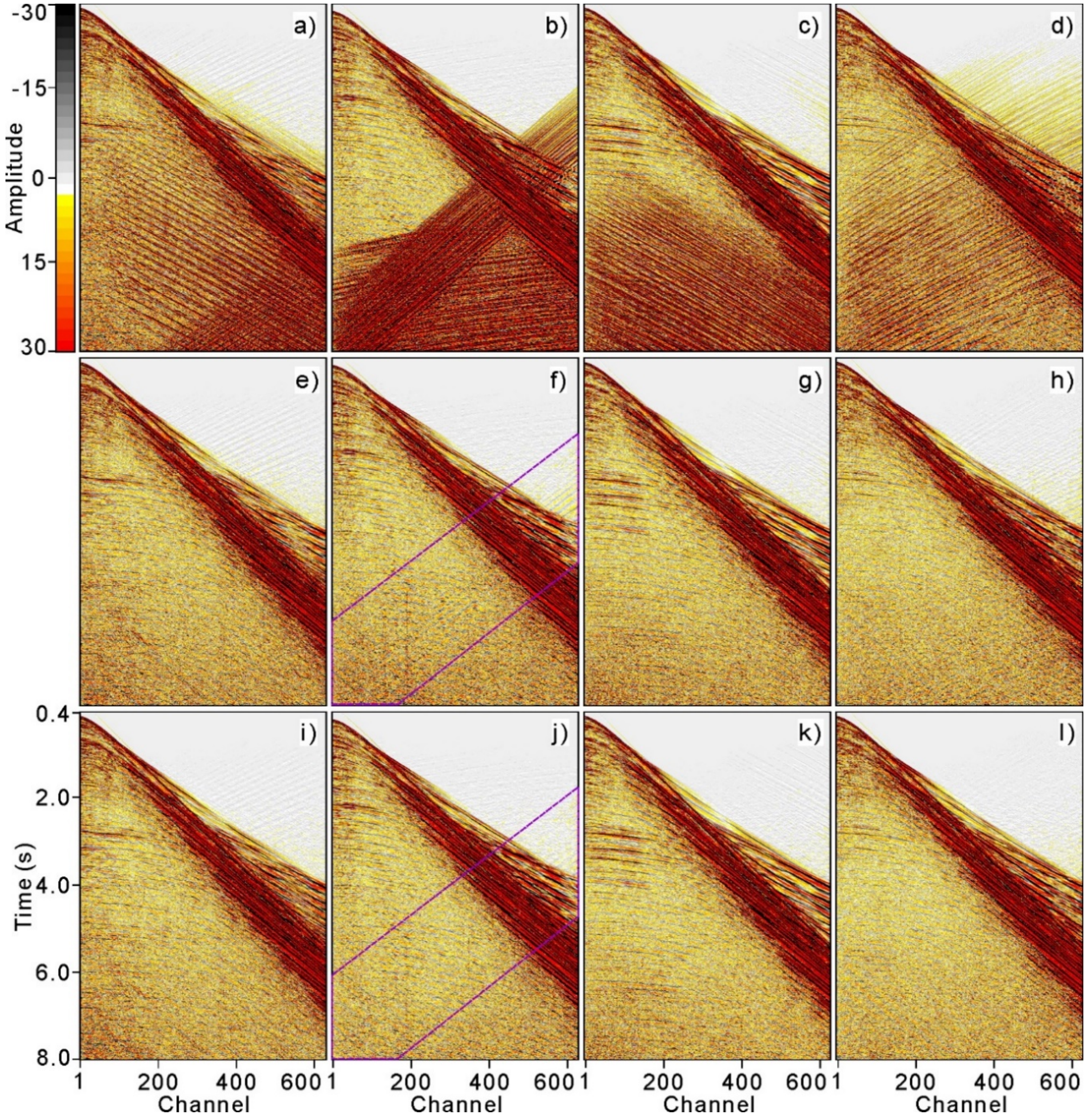

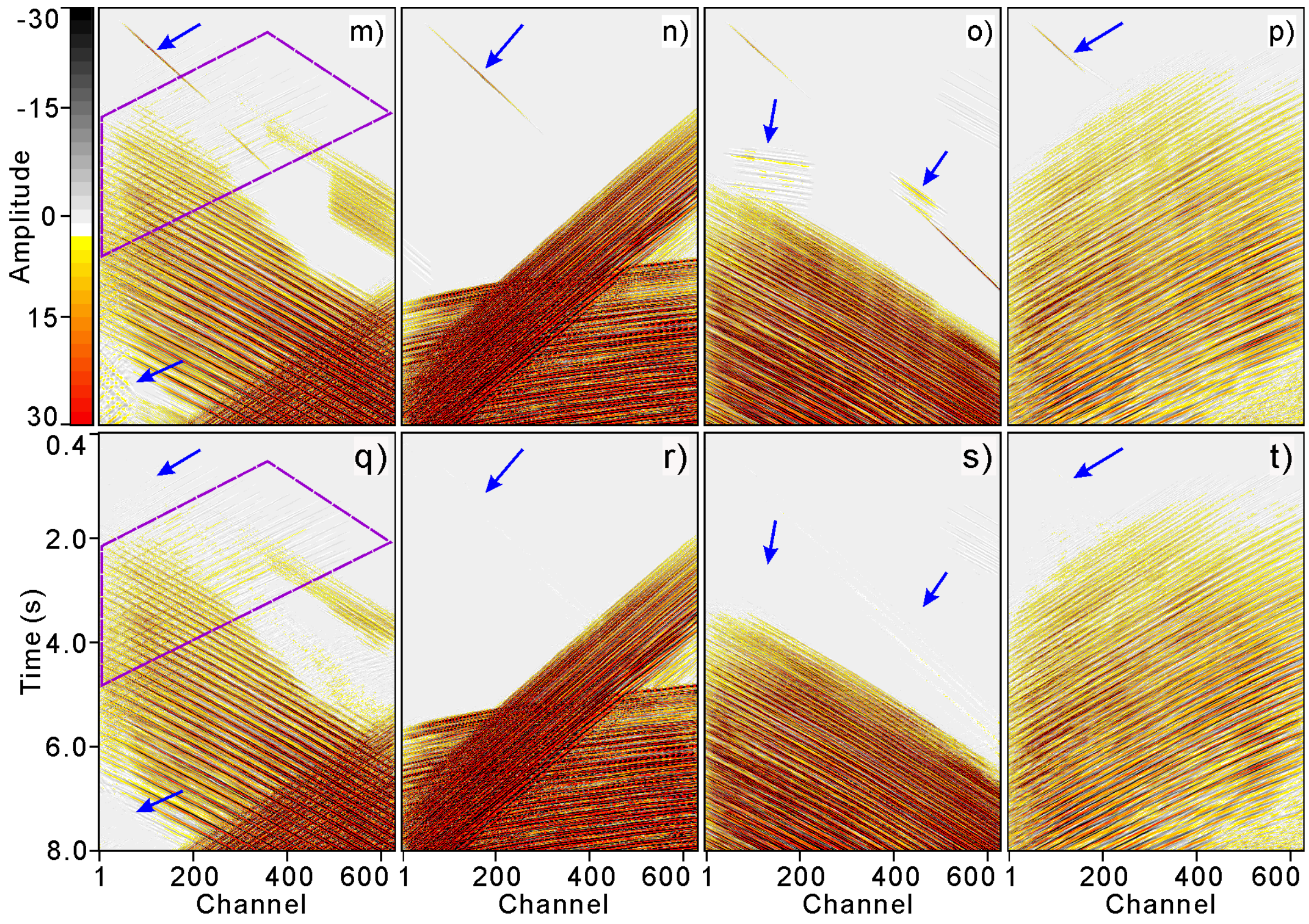

Figure 4: The first row (a) to (d) show four shots contaminated by different SI during the acquisition. The second row (e) to (h) show SI removal results using the conventional algorithm. The third row (i) to (l) show SI removal results using the proposed DNN-based workflow. The fourth row (m) to (p) show SI extracted by the conventional algorithm. The last row (q) to (t) shows SI predicted by the proposed DNN-based workflow.

**Data preparation**

As mentioned in the previous section, we built a training set in the following way. First, a small subset of the SI-contaminated data was passed through the above introduced conventional physics-based SI attenuation algorithm. In our case, around 5000 shots from three cables at different sail lines were selected (the location of the studied data in the whole NVG survey is marked by the green box in Figure 2), covering all the above-mentioned types of SI noise, which were also all the types of SI noise that the DNN would process at its application stage. Since in the studied area there are no SI-free CSGs from the acquisition, this conventional SI attenuation algorithm produced both the SI noise model and SI-free CSGs used in DNN training. Random scaling and shifting were applied to the direct outputs of the conventional algorithm so that more examples were obtained. Then, we randomly shuffled records from the SI noise model and blended them with the SI-free CSGs and a set of uniform random noise. Further, three adjacent shots on both sides of the to-be-processed shot were used as additional channels of the input. In the end, over 10,000 pairs consisting of a seven-channel input object and a two-channel desired output were obtained. Each shot gather used in this study has a recording length of 8 s two-way travel time with a 4 ms sample rate and 636 traces.

**DNN architecture and training**

The DNN we used is a U-Net (first proposed by Ronneberger et al., 2015). A visualization of its architecture can be found in Figure 5. The building block of the encoder consists of convolutional layers employing a typical filter size of 3×3 with the ReLU (Rectified Linear Unit) function and max pooling yielding a multilevel, multi-resolution feature representation. The max pooling employs a stride of 2 and a pool size of 2×2. The corresponding building block of the decoder employs transposed convolutional layers with a stride of 2 to up-sample low-resolution features. Skip connections between the encoding and decoding paths employ concatenation operations to ensure information fusion between high- and low-level information. No activation function is applied at the last layer. Adopting Adam optimization (Kingma and Ba, 2014), a batch size of 4 is selected. The DNN was trained to predict both the SI and a blend of the SI-free shot with the intact random noise, as described in the last section. 75% of the data pairs prepared above were used for the training while the remaining 25% were used for the validation of the DNN. The training and validation losses are plotted in Figure 6.

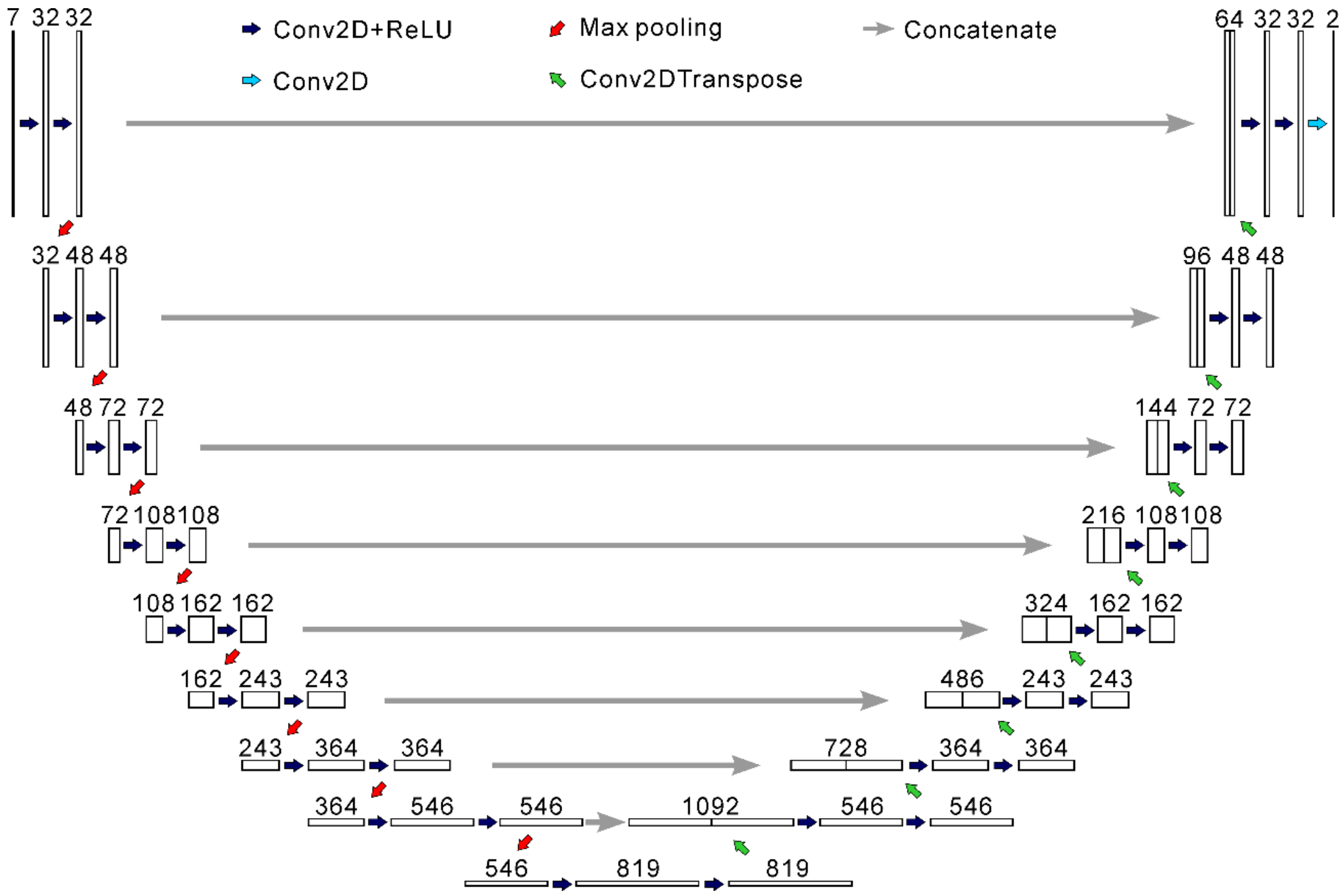

Figure 5: Architecture of the employed DNN. Each rectangle represents a collection of feature maps from the previous operation, and the numbers above represent the number of feature maps.

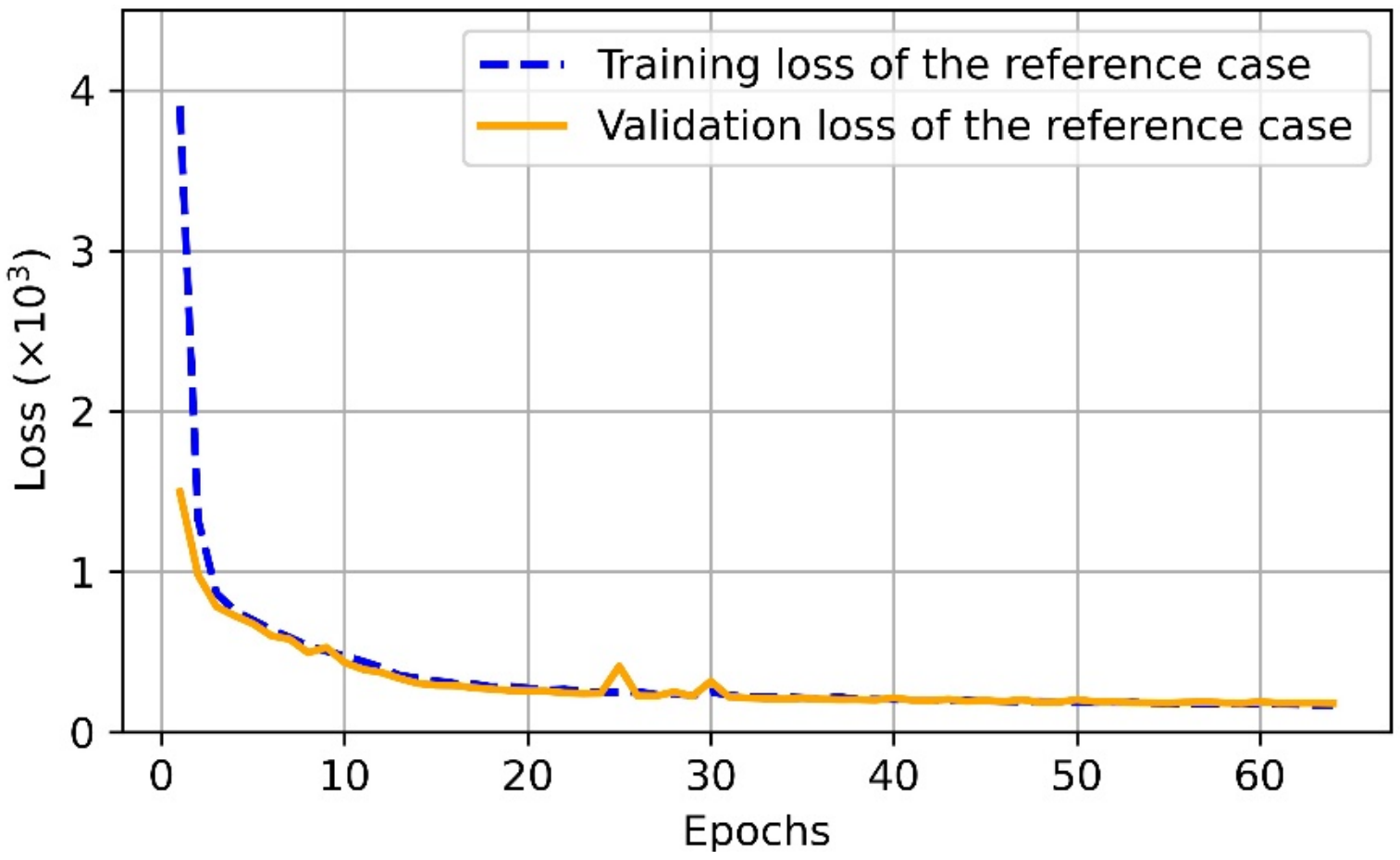

Figure 6: Plot of the training and validation losses.

**Results and comparison of the processing quality**

Figure 4 shows examples of the SI removal performance of the trained DNN on data unseen by the training (and validation) regime but from the covered sail lines. For comparison, we show corresponding examples from the conventional algorithm. Examples of four types of SI are shown in Figures 4a to 4d. Their corresponding SI removal results obtained by using the conventional algorithm are given in Figures 4e to 4h. The SI removal results obtained by using the proposed DNN-based workflow are shown in the row below (Figures 4i to 4l). Figures 4m to 4p and Figures 4q to 4t show SI removed using the conventional algorithm and the proposed DNN-based workflow, respectively. The purple boxes in Figures 4m and 4q show that the DNN-based workflow predicted more complete SI noise than the conventional workflow. Similarly, the SI noise residual in Figure 4j is less than that in Figure 4f. In addition, the blue arrows indicate where less signal leakage was caused by the DNN-based workflow compared to the conventional algorithm.

The deep RMS amplitude map is a commonly used QC metric for assessing the noise removal level in real processing projects of SI attenuation. Thus, the deep RMS levels of two cables from two sail lines are computed and displayed in Figure 7 to compare the field SI-contaminated data, the SI removal result using the conventional algorithm and the SI removal result using the proposed DNN-based workflow, which are the subfigures from top to bottom in each column. The RMS amplitudes are computed based on the average RMS per shot of late times (i.e., 6.0 s to 8.0 s), where most of the SI noise exists in the data of interest. As we can see, the proposed DNN-based workflow managed to remove more complete SI noise which appears predominantly as vertical stripes in these figures.

Furthermore, to better detect and assess the signal leakage, the data are resorted to the channel (offset) domain, with a group of examples given in Figure 8. Figure 8a shows an example of the field SI-contaminated data. The corresponding SI removal results from the conventional algorithm and the proposed DNN-based workflow are shown in Figures 8b and 8c respectively; the difference is difficult to see. A better comparison can be shown by the extracted SI from the conventional algorithm and the predicted SI from the proposed DNN-based workflow as shown in Figures 8c and 8e. As we observed in the shot-domain examples, the proposed DNN-based workflow outperformed the conventional algorithm in terms of processing quality, showing reduced signal leakage (indicated by the blue boxes).

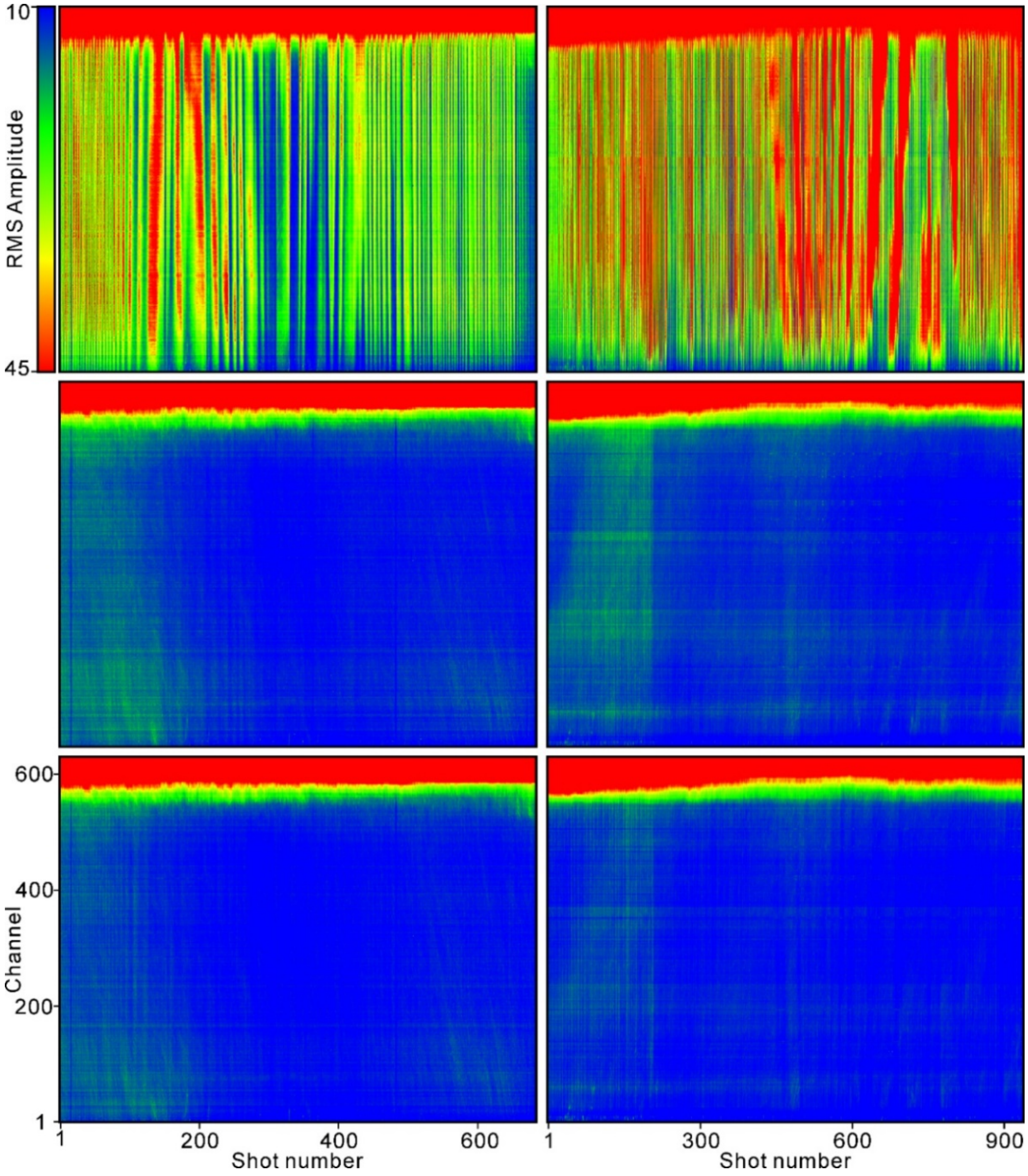

Figure 7: Deep RMS of two cables at different sail lines in the studied survey. In each column the figures from top to bottom are: SI-contaminated raw data, SI removal result using the conventional algorithm and SI removal result using the proposed DNN-based workflow.

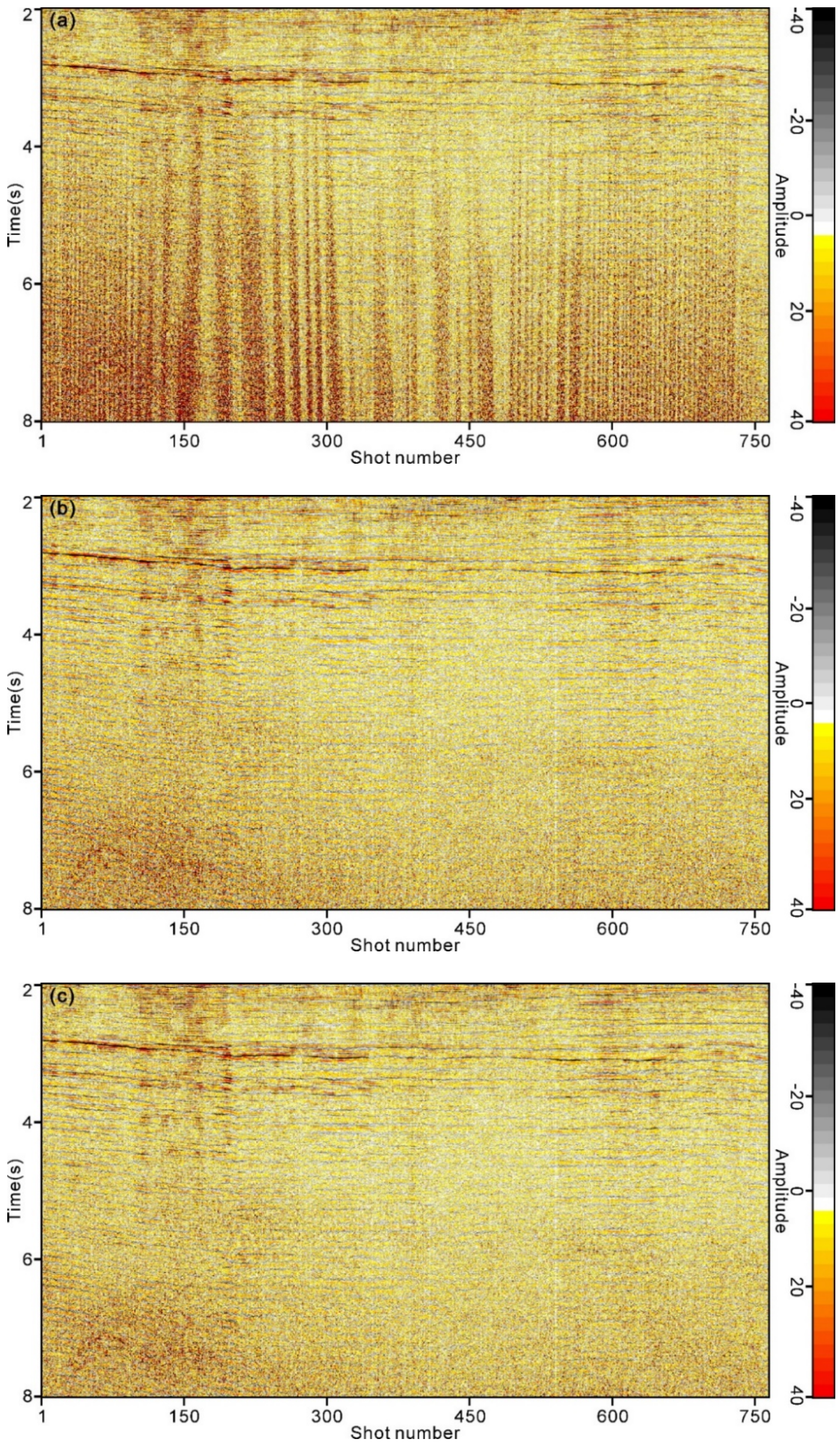

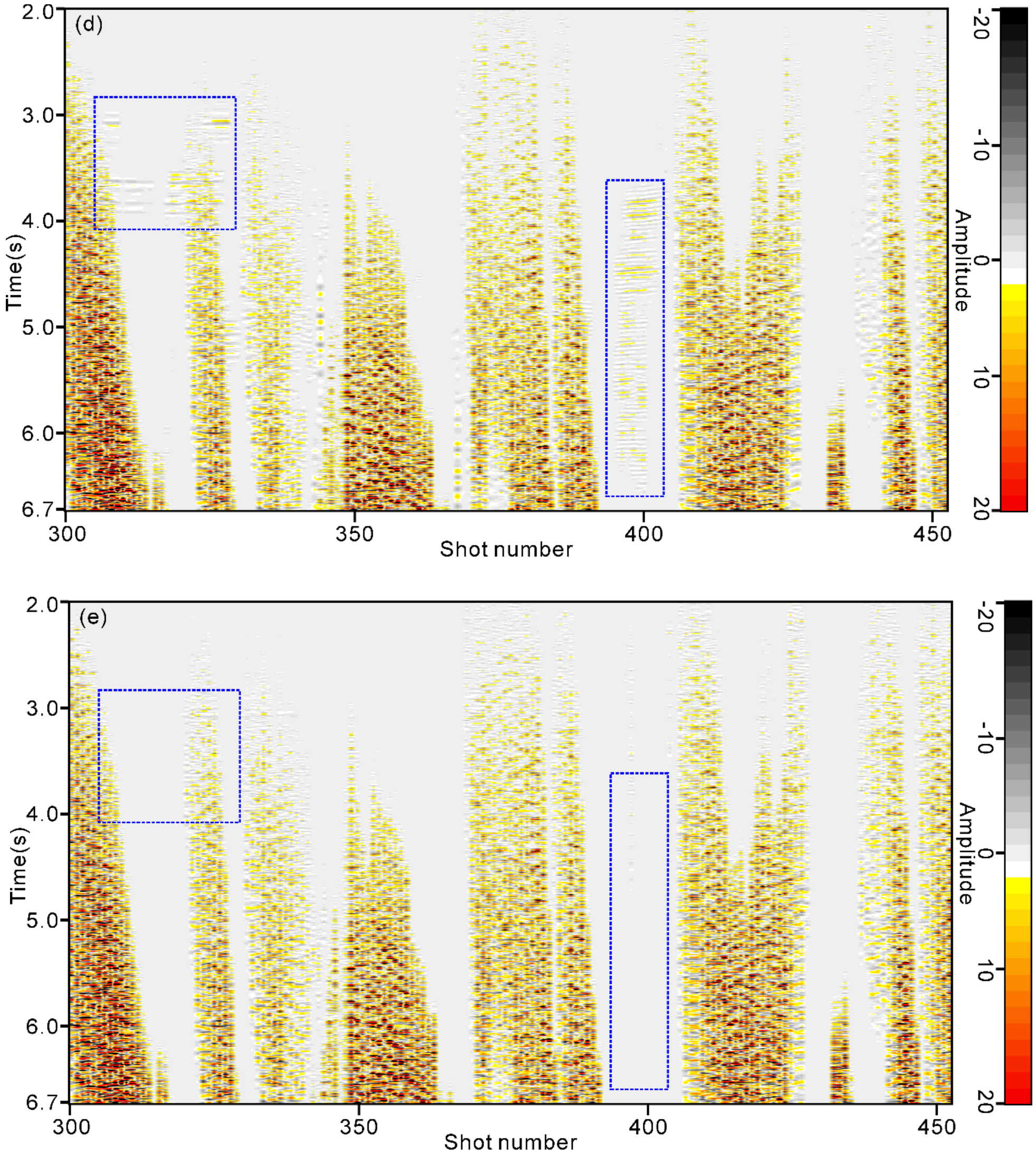

Figure 8: Examples in the channel (offset) domain of (a) the field data contaminated by SI during acquisition, (b) SI removal result using the conventional algorithm, (c) SI removal result using the proposed DNN-based workflow (d) extracted SI using the conventional algorithm and (e) predicted SI using the proposed DNN-based workflow.

**Computational efficiency comparison**

The processing time of this survey via conventional means varied significantly from sail line to sail line. This is because, despite being based on the same physical methodology, different workflows had to be used for different sail lines when their recorded SI noise differed significantly, and sometimes a second pass was required, in order to ensure a good SI attenuation result that could meet the production standard. In addition, it is not easy to quantitively measure the processing time of the conventional workflow since different processing units (central processing unit, abbreviated as CPU, or graphics processing unit, abbreviated as GPU) are used for different steps in the production line. In general, the runtime for different sail lines for the studied area varied from 18 to 36 hours per 3000 shot gathers, without considering the time of parameter optimization.

For the proposed DNN-based workflow, about 40 hours were spent getting the training and validation pairs for this study, including both the time periods of first sending a raw SI-contaminated data set through the conventional algorithm and then generating more examples through the subsequent preparation process. After that, the training process of the DNN took approximately 30 hours on the GPU, Nvidia Quadro RTX 6000. Once trained, the DNN could be applied to the rest of the data and could process each shot gather in approximately 0.1s. If the proposed DNN-based workflow was applied to a larger area, the training time of the DNN would increase only slightly, but the DNN's advantage in the application stage would be further magnified. Therefore, the overall relative time improvement of applying the proposed DNN-based workflow would scale up with the size of the processing area.

## DISCUSSION

Since 2018, DNN-related studies have emerged in the seismic community along with a stream of on-topic workshops and special sessions in the annual meetings of well-known learned societies in the geophysics community. The excitement wears off after a while, with DNNs' limitations and uncertainties springing to the surface, and more and more people are questioning DNN's production-oriented use in real seismic processing projects. Many studies have indicated that DNNs have an edge over elaborate but computationally heavy physics-based algorithms in terms of overall processing efficiency if the to-be-processed data set is large enough to tolerate the DNN's training phase; however, an outperformance in terms of processing quality is rarely seen. In this scenario, the observations in our study are clearly exciting and worthwhile to discuss.

For DNNs of a data-driven nature, the first issue to consider is always the training data. As has been demonstrated in many previous studies, for DNN with supervised learning, the quality of training data is critical regarding the effectiveness of the training process and the accuracy of the trained predictions (Sun et al., 2022); for example, if appreciable amplitude dimming and/or artefacts always appear in a certain section of the learning targets one after another, they will basically be inherited by the DNN into its future predictions after being trained. In this respect, despite not being perfect, the quality of the SI attenuation results from the conventional algorithm that we used to generate our DNN's training (and validation) pairs is very good. Although SI residual and signal leakage can be observed respectively in the SI-attenuated CSGs and SI noise models from the conventional algorithm, they are not significant nor are they regularly distributed in the whole data set.

Moreover, the list of tricks used in the proposed DNN-based workflow effectively mitigates the possible negative impact of such problems: The coherency of signal leakages among consecutive shot gathers in the SI noise models is broken by applying a random shuffle; using adjacent shot gathers as additional channels of the input reinforces the distinction between desired signal and SI noise; the employment of additional random noise in the training pairs enhances the DNN's learning attention to the high-level features of the data. With these efforts, the DNN would learn that those misallocated energies (SI residual in SI-attenuated CSGs and signal leakage in SI noise models) are undesired random interference to overlook rather than to mimic.

Secondly, the good aspects that DNNs show should not be ignored, even though they still cannot, and may never be able to, replace the physics-based methods. Conventionally, physics-based methods have no requirements on the amount of studied data and are of good interpretability and flexibility due to their basis in proven physical laws. They can deal with each given case in an isolated manner through adjustable parameters, sliding windows, etc. In contrast, a DNN approximates a function (a trained DNN model) for the whole data set of interest. That is to say, the DNN methods seek a global optimal solution for the collective, while the conventional physics-based methods seek the theoretically optimal solutions for each input (Hou et al., 2019; Hou and Messud, 2021). This enables a DNN to perform more robustly with data variations than a conventional physics-based method; for example, a trained DNN can deal with all four types of SI in one go. As a result, it is worthwhile to invest more in developing DNN-based technologies with the fair expectation of assisting human experts in labor-intensive tasks or supplementing already developed physical algorithms rather than overturning them.

## CONCLUSIONS

DNN-based methods have been considered useful but not widely deployed for real-world seismic processing projects, largely because their signal fidelity has not matched that of existing conventional algorithms. In this work, we have proposed a DNN-based workflow of real production value for separating SI noise from underlying reflection data on the CSGs. The proposed DNN-based workflow has been used for removing various types of SI noise in a field marine seismic survey conducted in the North Sea. We performed a comparison with an advanced conventional workflow (consisting of multiple steps) used in a real processing project. This conventional workflow was also used for providing the SI-free shots and SI noise model for the training and validation of the DNN.

The proposed DNN-based workflow shows an overall improvement over the conventional algorithm in terms of processing quality, particularly at handling SI noise with low amplitude in places with poor signal-to-noise ratio (S/N) or similar dips to the desired signal. The field data examples showed that the DNN-based workflow can perform effectively with SI noise being removed almost completely at the application stage, even though it was trained with the SI-free CSGs and SI noise model both produced from the conventional algorithm. Since the entire studied data set was seriously contaminated by SI noise during acquisition, the success of the proposed DNN-based workflow is very encouraging. However, we should realize that, for such a case, optimizing the conventional workflow is still important, as this is used to generate both the SI noise model and the SI-free shots for the DNN training and validation. A more ideal situation would be the presence of several field-acquired SI-free sail lines which we can use as ground truth to generate the DNN's training data. If this is the case, the requirement to take great care when applying the conventional algorithm might be relaxed, since only one (SI noise model) of the two ingredients (SI noise model and SI-free shots) is influenced. Overall, the success in processing this challenging survey validates the value of the proposed DNN-based workflow for real-world processing projects. Moreover, although this DNN-based workflow was proposed for the separation of seismic signal and SI noise, it has the potential to be adapted to the removal of other types of coherent noise.